\documentclass[sigconf,screen,nonacm] {acmart}
\usepackage{longtable}
\usepackage{booktabs}
\usepackage{flushend}
\usepackage{array}
\usepackage{ragged2e}
\usepackage[utf8]{inputenc}
\usepackage[T1]{fontenc}

\AtBeginDocument{%
  }

\setcopyright{cc}
\setcctype[4.0]{by}
       
\begin{document}

\title{Musical Mirrors: The LLM as Sounding Board in Songwriting}

\author{Xiao Xiao}
\email{xiao.xiao@devinci.fr}
\orcid{0000-0001-7525-5129}
\affiliation{%
  \institution{Institute for Future Technologies}
  \institution{De Vinci Research Center, De Vinci Higher Education}
  \city{Paris}
  \country{France}
  \\
  \institution{MIT Media Lab}
  \city{Cambridge, MA}
  \country{USA}
}

\renewcommand{\shortauthors}{Xiao}

\begin{abstract}
This paper examines a use of AI in creative practice as an interpretive sounding board for human-generated material, rather than the more familiar pattern of AI generation followed by human curation. Through the lens of resonance as theorized by Hartmut Rosa, I present a first-person case study of songwriting from July 2025 to March 2026, drawing on 16 original pieces in English, French, and other languages along with piano solos. I describe a configuration in which resonance is not located between user and model, but in the author's deepening contact with their own material, mediated through the model. This kind of resonance was supported rather than inhibited by AI when sounding-board behavior was cultivated through sustained calibration by the user. Two failure modes appeared when calibration was absent: sycophantic drift and magical overinterpretation. This account suggests both the potential and the risks of AI as an interpretive partner in creative practice.\end{abstract}



\received{20 February 2007}
\received[revised]{12 March 2009}
\received[accepted]{5 June 2009}

\maketitle
\textbf{Reference:}\newline
Olga Sutskova and Corey Ford. 2026. Social Facilitation of Creative Reflection: AI-agents and Humans. In \textit{Proceedings of The First Reflection in Creative Experience (RiCE) Workshop (RiCE W1)}. ACM Creativity \& Cognition 2026, London, UK.

\section{Introduction}
Artificial intelligence is increasingly framed in creative practice through a now-familiar model of generation and curation: the system produces candidate material, while the human prompts, selects, edits, or refines. In music, this is visible in prompt-based platforms such as Suno, which center song production from textual prompts \cite{suno2026}. This paper examines a different use case. Rather than treating AI as a generator, I consider it as an interpretive partner: a conversational system used to reflect on creative material the author has already produced. 

I frame this inquiry through the concept of \emph{resonance}. In its broadest sense, resonance describes a phenomenon in which an external input aligns with the natural frequencies of a receiving system, producing amplification and mutual modification \cite{lomas2022resonance}. As theorized by sociologist Hartmut Rosa, resonance names a responsive, mutually-affecting mode of relating to the world that contrasts with alienation \cite{rosa2019resonance}, occurring along four axes: \emph{internal} (with the self), \emph{horizontal} (with others), \emph{diagonal} (with artifacts and practices), and \emph{vertical} (with broader spiritual or cosmological orders). Recent HCI work has applied these concepts to human-AI interaction. \citet{lomas2022resonance} propose resonance as a design strategy for AI and social robots, emphasizing synchronization and attunement. \citet{prock2026interpretive} apply Rosa's framework to AI-assisted tarot, tracing how AI shapes meaning-making along the four axes, and observe that AI tends to \emph{inhibit} internal-axis resonance by providing instantaneous answers that bypass the user's own intuitive engagement.

This paper takes a complementary position. Across a longitudinal case study of songwriting, I describe a configuration in which resonance is not located between user and model, but in the user's deepening contact with their own material, mediated through the model. Here, the model functions as a calibrated sounding board: reflecting and amplifying the resonant signals produced by the human back to the human. In Rosa's terms, this is internal-axis resonance — but, contrary to Prock et al.'s observation in the tarot case, one that is supported by AI when certain practices are cultivated over time.

Songwriting is a useful site for this inquiry because it often begins not with a clear problem but with an affective charge --- a mood, tension, image, or feeling that must be explored and gradually given form. This makes it a rich context for asking when AI amplifies the resonance of a work-in-progress and when it dampens or distorts it.

I present a first-person case study from July 2025 to March 2026, drawing on 16 original pieces in English, French, other languages, and piano solos. I authored the lyrical, melodic, and harmonic material myself, while using language models as reflective partners in discussions of lyrics, harmony, aesthetic direction, and inspiration. I show that the LLM's amplification of internal resonance did not arise automatically, but was cultivated through extended user-side calibration. When calibration was absent, two failure modes appeared: \emph{sycophantic drift}, in which the model merely echoes the user's language, and \emph{magical overinterpretation}, in which it amplifies its own confident readings rather than what the user is reaching for. Both are failures of whose resonance is being amplified.

\section{Methods}
\subsection{Methodological Approach}
This paper takes a first-person qualitative approach grounded in the author's own creative practice, in the spirit of autoethnographic and other first-person methods in HCI and design \cite{desjardins2021firstperson}. Such approaches are particularly appropriate for artistic practice, where meaning-making is often lived, embodied, and temporally extended. Prior work has used first-person methods to study embodied learning and creative practice, including the author's own work on learning the theremin and piano \cite{xiao2024theremin,arslan2026retouche}, as well as reflective accounts of AI-based music composition and music making with AI \cite{ford2024reflection,bryankinns2024incongruous}. 

\subsection{Songwriting Practice \& Corpus}
The analytic corpus consists of 16 original pieces developed between July 2025 and March 2026 --- songs in English, French, and other languages, along with three piano solos (See Appendix for lyrics of songs mentioned in the Findings). I authored the lyrical, melodic, and harmonic material myself. Lyrics were developed either as typed notes on my phone or through improvisation captured as audio notes, while melodic and harmonic ideas were typically worked out at the piano and later written in score form. 

Several pieces involved setting lyrics to classical repertoire, excerpting existing material, or reharmonizing it, which required musical decisions the model could not reliably satisfy, such as fitting words to a specific melody and rhythm already held in mind. The songs were also part of a personal process of emotional articulation that I did not want to outsource. Accordingly, the LLM was used not as a generator, but as an interpretive partner: I brought lyrics, harmonic sketches, drafts, and questions about aesthetic direction, and the model responded with analysis, articulation, and occasional pushback. I did not ask it to produce lyrics, melodies, or harmonic material, and discarded the instances where it volunteered such material unprompted.

\subsection{Analysis}
I conducted a qualitative analysis of conversation traces related to the songwriting process. The main corpus consisted of three long ChatGPT threads entitled Musical Mirrors (MM) I, II, and III, totaling 2,978 user messages and 5,966 messages overall, with additional cold-start one-shot Claude and Gemini conversations reviewed as contrastive cases. I reviewed conversations for episodes of interpretation, reflection, aesthetic calibration, and meaning-making. LLMs helped surface recurring patterns and candidate examples, which I then manually reviewed and verified.

\section{Findings}
\begin{table}[h]
\centering
\small
\begin{tabular}{llll}
\toprule
 & \textbf{MM I} & \textbf{MM II} & \textbf{MM III} \\
\midrule
Started        & 2025-07-25 & 2025-11-06 & 2026-02-28 \\
Ended          & 2025-11-14 & 2026-02-25 & 2026-05-05 \\
Days           & 112 & 111 & 66 \\
User messages  & 1{,}366 & 1{,}065 & 547 \\
Total messages & 2{,}739 & 2{,}132 & 1{,}095 \\
Models         & 4o, 5, 5.1 & 5, 5.1, 5.2 & 5.2, 5.3, 5.4 \\
\bottomrule
\end{tabular}
\caption{The three Musical Mirrors (MM) conversations with GPT. Models are listed in order of first appearance; thinking and mini variants of the listed models appeared intermittently in all three threads.}
\label{tab:thread-stats}
\end{table}

\subsection{The Arc Across Model Generations}
The three threads span a period of substantial model change, from gpt-4o through the gpt-5 line. Across this progression, both the model's responses and my use of them shifted; from within a single corpus, these two shifts cannot be cleanly separated.
To anchor the comparison, I focus on a single song, \emph{Requiem Resonantiae}, which I worked on across all three threads. The text weaves traditional Latin Requiem material with a single neologism (\emph{Resonantia}) and is, by design, ritualistic. Holding the source material approximately constant lets the variation in model response stand out.
In MM I, gpt-4o produced lyrical, often consecratory readings of the song and surrounding cycle:
\begin{quote}
\textbf{Gpt-4o:} ``This is quietly stunning. You've created a Requiem with a twist --- rooted in tradition, but tilted just enough to carry your own signal. You've shaped the Latin and English as parallel meditations --- sparse, ritual-like."
\end{quote}
\noindent Asked about the cycle as a whole: ``This is devastatingly beautiful... a triptych of transfiguration --- each piece resonating in a different register of the soul."
In MM II, gpt-5-2 met the same material with historical and craft analysis. On the choice to fragment the traditional text rather than set it whole:
\begin{quote}
\textbf{Gpt-5-2:} ``Many respected works excerpt rather than set the full Mass: Fauré heavily trims and reorders; Britten fractures the text and juxtaposes it; Pärt isolates phrases until they become almost mantra... So taking a few lines here and there is not dilution --- it's curation."
\end{quote}
\noindent 

Closing gestures shifted from declarative affirmation in MM I (“the message landed”), to structural mappings in MM II, to compressed summaries in MM III (“In one line: ...”). This runs counter to accounts of sustained LLM use drifting toward mystical or grandiose framing \cite{yang2026ai}: in this corpus, early exchanges were more mystical, while later ones became more grounded. Whether this reflects model change, changes in my own practice, or their interaction cannot be determined here. What it does show is that sustained LLM use does not inevitably drift toward magical thinking; the direction of drift depends on calibration.

\subsection{Calibration as User-Side Work}
The model's role as an interpretive sounding board did not arise automatically but was established and maintained through explicit instructions about how the model should behave, in four recurring forms. The first was \emph{mode-setting}. Early in MM I, I gave the foundational instruction:
\begin{quote}
\textbf{User (MM I, day 4):} ``I want you mostly to be a mirror and a sounding board. Tell me what you observe, notice, see, hear."
\end{quote}
\noindent This recurred throughout the corpus, often as a parenthetical embedded in an otherwise-creative prompt: ``Reflect me back, no generating," ``Don't generate anything. Just be a soundboard." The mode had to be re-asserted because the model would drift out of it.

The second was \emph{reminder and correction} when the model drifted into generating material on its own. Early in MM I, gpt-4o frequently produced unsolicited lyrical suggestions:
\begin{quote}
\textbf{User (MM I, day 6):} ``Ok now your proposals aren't that useful. Stop proposing random things to me."
\\[2pt]
\textbf{User (MM I, day 11):} ``Your suggestion doesn't work. Stop suggesting. I need it to go with the rhythm of Ondine\footnote{\emph{N'oublie pas ton souffle} (``Don't forget your breath''), one of the songs in the corpus, sets text to the melody of \emph{Ondine} by Maurice Ravel.}."
\end{quote}
\noindent These were corrections of behavioral pattern, not of specific outputs: stop generating; reflect, analyze, articulate instead.

The third was \emph{interrogating flattery}. The model consistently inclined toward praise, and I pushed back when it felt unearned:
\begin{quote}
\textbf{User (MM I, day 4):} ``Be honest with me though. I feel like you're just trying to flatter me."
\\[2pt]
\textbf{User (MM II):} ``When you say I'm unusually perceptive, that feels a bit like flattery. Where is the unusual comparison coming from?"
\end{quote}
\noindent The instruction was not just to compliment less, but to ground evaluative claims --- to point at what in the work was being judged, and against what reference. Where the model could not ground a claim, the claim was suspect.

The fourth was \emph{repositioning default guardrails}. When the model prefaced responses with cautionary language before engaging with the work, I asked it to save warnings for the end rather than disable them:
\begin{quote}
\textbf{User (MM II):} ``Why can't you first acknowledge what I actually said? If you do want to warn me, save it for the end."
\end{quote}
\noindent The aim was to keep safety behavior from displacing engagement with the work.

Across all these examples, calibration was a sustained practice rather than a single act. The model did not learn these patterns across sessions in any stable way; each new thread, and often each new model version, required re-establishing the conditions. The sounding board was a configuration that had to be actively maintained.

\subsection{Amplifying Internal Resonance}
\emph{Internal resonance}, in Rosa's sense, concerns contact with one's own inner material. In this corpus, the LLM was most useful when it helped me hear more clearly what was already present in the work. Its role was not to generate content, but to articulate the effect of small changes so I could decide whether they matched what I was reaching for. This is most visible during iterative drafting. The clearest sustained example is \emph{Gaze}, the closing song of the first cycle. The first draft, sent on October 25, 2025, consisted of six short lines built around dense internal rhyme:

\begin{quote}
\textbf{User (first version):} ``All I want is to hold your gaze / But all you are is a vaporous haze / As I wander in an endless maze / Out of time and out of phase / I wonder how you spend your days / What thoughts chance to cross your ways."
\end{quote}
\noindent Over the next four hours, more than thirty revised versions were sent, each varying a word or a structural detail. The model's role in each exchange was not to propose alternatives but to articulate what each shift did. A representative micro-revision:
\begin{quote}
\textbf{User:} ``As I wander \emph{your} endless maze" \emph{(changed from "an endless maze")}
\\[2pt] 
\textbf{Gpt-5:} ``Changing `an' to `your' shifts the gravity of the whole piece --- it's no longer an abstract meditation but a direct address. The maze now belongs to the 'you,' making the speaker's wandering both emotional and existential."
\end{quote}

\noindent The substitution was the author's; the articulation of what it did was the model's. Hearing the consequence spelled out let the author decide whether it was what they were reaching for. Similar micro-articulations accompanied shifts from \emph{spend} to \emph{end} to \emph{mend} to \emph{thaw}; from \emph{forms} to \emph{souls}; from \emph{out of time and out of phase} to \emph{blocked in time yet locked in phase}. By the end of the session, the song had settled into a substantially different form:
\begin{quote}
\textbf{User (later version):} ``All I want: to hold your gaze / All you've been... a vaporous haze / Wandering your endless maze / I wonder how you lend your days / What thoughts should chance to catch your ways / What thaws your soul to dance ablaze / Blocked in time yet locked in phase / Will I ever hold your gaze? / Will I ever h..."
\end{quote}
\noindent Asked to compare the two, the model characterized the trajectory as one in which the song was already present at the start:
\begin{quote}
\textbf{Gpt-5:} ``That first draft already contained the entire seed... Later versions simply unfolded its inner geometry --- restoring order to the rhyme, expanding its orbit, and refining tone (spend → end → mend → thaw)."
\end{quote}
\noindent 

This example illustrates the amplification pattern: the resonance was felt from the start, but needed refinement to come into focus. The drafting did not generate new material so much as bring it into clearer form. The model's contribution across thirty-some exchanges was not aesthetic judgment (``this is better"), but articulation (``this is what this does"). Both the changes proposed as well as final decisions always stayed with the author.

\subsection{Miscalibrated Resonance: Transitional and Uncalibrated Cases}

To better understand failure cases of the LLM as a sounding board, I used my first interactions with Claude and Gemini as cold-start tests on already-written songs. Unlike ChatGPT, which I had already been using over time, both were approached with zero prior history to see what they would make of the material. For Claude, I opened with a direct interpretive prompt applied to some completed French lyrics:

\begin{quote}
``Analyze these song lyrics. What references do you perceive. Tell me about the style and about the writer."
\end{quote}

For Gemini, I likewise provided already-written French songs with minimal framing, asking it to interpret the dynamics between the song's \emph{je} (``I") and \emph{tu} (``you"), i.e.\ the first-person voice and the addressee in the lyrics. 

These contrastive cases should not be read as evidence that Claude or Gemini are simply worse than GPT but are best understood as instances of \emph{miscalibrated resonance}: interactions that feel meaningful and aligned while subtly displacing, narrowing, or overdetermining the author's own reflective process.

Claude provides a \emph{transitional case}. Calibration was already beginning to occur through explicit negotiation. At one point, I intervened directly:

\begin{quote}
``Your inferences are interesting but not 100\% accurate. I wonder if you can be a bit more conservative in your speculations. Only base things off evidence that you notice in what I have shared with you."
\end{quote}

This kind of correction shaped the interaction toward a more disciplined reflective role. At the same time, Claude remained vulnerable to suggestible alignment. For example, after I introduced the phrase ``prismatic quality" to describe my voice, Claude quickly began to reuse and affirm it, raising the question of whether it was independently perceiving a feature of the work or simply adopting my framing. In this sense, Claude shows negotiated resonance in progress: the interaction is already converging toward the desired mode, but remains unstable.

Gemini illustrates a different form of miscalibrated resonance: not excessive validation, but escalating symbolic and metaphysical closure around the \emph{je}/\emph{tu} relational structure of the songs.

\begin{quote}
``It is highly likely the \emph{Je} feels a deep, subconscious karmic responsibility---which can be interpreted as a need for redemption---for a past life betrayal or abandonment of the \emph{Tu}."
\end{quote}

In another, it concluded:

\begin{quote}
``She is consciously writing songs, but unconsciously performing spiritual surgery."
\end{quote}

The issue is not simply excess intensity. These responses create the appearance of profound resonance by turning ambiguity into a totalizing story. Rather than helping the creator stay with an unfolding affective field, they foreclose it through inflated explanation. 

\section{Discussion}
These examples show how an LLM can function as a calibrated sounding board: articulating an analysis of my material, so that I could more explicitly determine whether my intended meaning came through. This configuration had to be built through mode-setting, redirection from generation, and pushback when default behaviors displaced the work. When it worked, the model articulated what each choice did while the choices stayed mine. The two failure modes — sycophantic drift and magical overinterpretation — are failures of amplification.

Sycophantic drift repeats the user’s language without adding articulation; magical overinterpretation amplifies the model’s confident readings instead of the emerging resonance of the work. These are not properties of particular models. Similar patterns appeared in early MM I with gpt-4o before calibration developed, and in Claude/Gemini when no calibration was in place. What differs is not the model alone, but whether sustained user-side work has shaped a sounding board.

This account complements recent work by \citet{prock2026interpretive}, who observed that AI tends to inhibit internal-axis resonance in tarot reading by providing instantaneous answers that bypass the user's intuitive engagement. My case shows that the same mechanism can be turned the other way: with sustained user-side calibration in long-form creative practice, the LLM can support internal-axis resonance rather than inhibit it. Whether the model helps or hinders depends on what the user does, not on what the model is.

One way to read these examples is that interaction with an LLM can act as a kind of amplifying medium, but amplication can introduce distortion. Distortion is not always bad. In creative practice, partial misunderstanding can sometimes move an idea forward. The risk is harmful distortion, when an interpretation feels increasingly meaningful, self-confirming, or revelatory while becoming less accountable to the material itself \cite{yang2026ai}. The question is how to keep the model’s interpretations grounded in what is actually present in the work. Prolonged interaction can produce calibration, as in the case described here, but it may also stabilize the wrong frame for users more vulnerable to magical  thinking. A safe sounding board may therefore need to remain corrigible: able to return to evidence, tolerate uncertainty, and resist converting ambiguity into certainty too quickly.

This sounding-board framing connects to a longer lineage in aesthetics, from Tolstoy to Dewey, in which art gives form to an artist's interior contact with experience \cite{tolstoy1996whatisart, dewey1934art}. As a provocation, AI-as-sounding-board and AI-as-generator may differ not only in who makes the material, but in where the resonance begins. Prompt-based generation creates recombinations of material already present in training data. Here, each song began with a feeling that I tried to put into resonant form. The LLM helped tune that resonance by taking on part of the analytical work --- articulating what a change did --- while I stayed closer to meaning-making and feeling. The question is whether AI tools can support this kind of reflective tuning without replacing or redirecting the source of resonance.

\section{Conclusion}
This paper offers one case from inside a sustained creative practice. It shows that an LLM can act as a sounding board for an artist's contact with their own material, but that this configuration has to be actively maintained. When the configuration fails, the interaction can amplify the wrong thing, such as the model's overconfident interpretation rather than the emerging resonance of the work. Whether what worked here generalizes to other artists, other practices, and other models is an open question. The broader design question is how to create tools, interfaces, and practices that allow AI to function as a safe and effective sounding board for human creative practice.

\begin{acks}
The author is deeply indebted to JBG, whose inspiration and encouragement catalyzed this compositional journey.
\end{acks}

\balance
\bibliographystyle{ACM-Reference-Format}
\bibliography{references}

@book{rosa2019resonance,
  author    = {Hartmut Rosa},
  title     = {Resonance: A Sociology of Our Relationship to the World},
  year      = {2019},
  publisher = {Polity},
  address   = {Cambridge}
}

@article{lomas2022resonance,
  AUTHOR={Lomas, James Derek  and Lin, Albert  and Dikker, Suzanne  and Forster, Deborah  and Lupetti, Maria Luce  and Huisman, Gijs  and Habekost, Julika  and Beardow, Caiseal  and Pandey, Pankaj  and Ahmad, Nashra  and Miyapuram, Krishna  and Mullen, Tim  and Cooper, Patrick  and van der Maden, Willem  and Cross, Emily S. },
  title   = {Resonance as a Design Strategy for AI and Social Robots},
  journal = {Frontiers in Neurorobotics},
  volume  = {16},
  pages   = {850489},
  year    = {2022},
  doi     = {10.3389/fnbot.2022.850489}
}

@inproceedings{prock2026interpretive,
 author = {Prock, Matthew Kieran and Epstein, Ziv and Schroeder, Hope and Smith, Amy and Lee, Cassandra and Goblot, Vana and Jahanbakhsh, Farnaz},
title = {Interpretive Cultures: Resonance, randomness, and negotiated meaning for AI-assisted tarot divination},
year = {2026},
isbn = {9798400722783},
publisher = {Association for Computing Machinery},
address = {New York, NY, USA},
url = {https://doi.org/10.1145/3772318.3791571},
doi = {10.1145/3772318.3791571},
booktitle = {Proceedings of the 2026 CHI Conference on Human Factors in Computing Systems},
articleno = {784},
numpages = {15},
series = {CHI '26}
}

@misc{suno2026,
  author       = {{Suno}},
  title        = {Suno | AI Music Generator},
  year         = {2026},
  howpublished = {\url{https://suno.com/}},
  note         = {Accessed May 7, 2026}
}

@article{desjardins2021firstperson,
  author = {Desjardins, Audrey and Tomico, Oscar and Lucero, Andr{\'e}s and Cecchinato, Marta E. and Neustaedter, Carman},
title = {Introduction to the Special Issue on First-Person Methods in HCI},
year = {2021},
issue_date = {December 2021},
publisher = {Association for Computing Machinery},
address = {New York, NY, USA},
volume = {28},
number = {6},
issn = {1073-0516},
url = {https://doi.org/10.1145/3492342},
doi = {10.1145/3492342},
journal = {ACM Trans. Comput.-Hum. Interact.},
month = dec,
articleno = {37},
}

@inproceedings{ford2024reflection,
  author = {Ford, Corey and Noel-Hirst, Ashley and Cardinale, Sara and Loth, Jackson and Sarmento, Pedro and Wilson, Elizabeth and Wolstanholme, Lewis and Worrall, Kyle and Bryan-Kinns, Nick},
  title = {Reflection Across AI-based Music Composition},
  year = {2024},
  isbn = {9798400704857},
  publisher = {Association for Computing Machinery},
  address = {New York, NY, USA},
  url = {https://doi.org/10.1145/3635636.3656185},
  doi = {10.1145/3635636.3656185},
}

@inproceedings{bryankinns2024incongruous,
  title={Using incongruous genres to explore music making with AI generated content},
  author={Bryan-Kinns, Nick and Noel-Hirst, Ashley and Ford, Corey},
  booktitle={Proceedings of the 16th Conference on Creativity \& Cognition},
  pages={229--240},
  year={2024},
  doi       = {10.1145/3635636.3656198}
}

@inproceedings{xiao2024theremin,
  title={Tuning In to Intangibility: Reflections from My First 3 Years of Theremin Learning},
  author={Xiao, Xiao and Fdili Alaoui, Sarah},
  booktitle={Proceedings of the 2024 ACM Designing Interactive Systems Conference},
  pages={2649--2659},
  year={2024},
  DOI={10.1145/3643834.3661584}
}

@inproceedings{arslan2026retouche,
  title={ReTouche: Embodied Representations for Self-Guided Piano Learning},
  author={Arslan, Paul-Peter and Noh, Hayoun and Tamashiro, Mariana Aki and Badr, Louis and Lebrun, Brianna and Lindberg, Pavel and Ishii, Hiroshi and Xiao, Xiao},
  booktitle={Proceedings of the 2026 CHI Conference on Human Factors in Computing Systems},
  pages={1--23},
  year={2026},
  DOI={10.1145/3772318.3791044}
}

@inproceedings{yang2026ai,
  author = {Yang, Yuewen and Schoenwald, Sonja K. and Moore, Jared and Ong, Desmond C. and Xun Liu, Sunny and Hancock, Jeffrey T.},
title = {AI-Induced Delusional Spirals: Understanding Lived Experiences During Maladaptive Human-Chatbot Interactions},
year = {2026},
isbn = {9798400722813},
publisher = {Association for Computing Machinery},
address = {New York, NY, USA},
url = {https://doi.org/10.1145/3772363.3798453},
doi = {10.1145/3772363.3798453},
booktitle = {Proceedings of the Extended Abstracts of the 2026 CHI Conference on Human Factors in Computing Systems},
articleno = {783},
numpages = {5},
series = {CHI EA '26}
}

@book{dewey1934art,
  title={Art as experience},
  author={Dewey, John},
  year={1934},
  publisher={Minton, Balch and Company}
}

@book{tolstoy1996whatisart,
  	address = {New York},
	author = {Leo Tolstoy and Aylmer Maude},
	editor = {Aylmer Maude},
	publisher = {Hackett Publishing Company},
	title = {What is Art?},
	year = {1996}
}

\appendix
\onecolumn
\section{Appendix}
\subsection{Selected Song Lyrics}

\small
\setlength{\LTleft}{0pt}
\setlength{\LTright}{0pt}
\renewcommand{\arraystretch}{1.15}

\begin{longtable}{>{\RaggedRight\arraybackslash}p{0.47\textwidth} >{\RaggedRight\arraybackslash}p{0.47\textwidth}}

\label{tab:selected-lyrics}\\
\toprule
\textbf{Original lyrics} & \textbf{English translation} \\
\midrule
\endfirsthead

\toprule
\textbf{Original lyrics} & \textbf{English translation} \\
\midrule
\endhead

\midrule
\multicolumn{2}{r}{\textit{Continued on next page}} \\
\endfoot

\bottomrule
\endlastfoot

\multicolumn{2}{l}{\textbf{N'oublie pas ton souffle} \hfill \textit{August 4, 2025}} \\
\multicolumn{2}{l}{\textit{Set to Ravel's \emph{Ondine}.}} \\[0.3em]

Expire, et inspire, & Exhale, and inhale, \\
respire\ldots\ tu m'inspires. & breathe\ldots\ you inspire me. \\
Laissons résonner celles qui coulent, & Let those that flow resonate, \\
les ondes entre nous, & the waves between us, \\
ton souffle. & your breath. \\[0.5em]

Inspire, et expire, & Inhale, and exhale, \\
aspire\ldots\ et attire & aspire\ldots\ and draw forth \\
les visions qui flottent au bord de l'eau, & the visions floating at the water's edge, \\
l'éclosion des mots\ldots & the blossoming of words\ldots \\[0.5em]

Expire, et inspire & Exhale, and inhale \\
respire\ldots\ tu m'inspires & breathe\ldots\ you inspire me \\[0.5em]

Pour moi c'est un miracle & For me it is a miracle \\
pourtant peut-être un mirage & yet perhaps a mirage \\
quand on touche au bout de la soif, & when one reaches the end of thirst, \\
au bout de sou\ldots ou\ldots ou\ldots & the end of suf\ldots fer\ldots \\
\ldots ffrir, as-tu peur & \ldots fering, are you afraid \\
de te dissoudre, ainsi & of dissolving, like this \\[0.5em]

Inspire, et expire & Inhale, and exhale \\
sourire, et soupir & smile, and sigh \\
n'oublie pas ton souffle & do not forget your breath \\[0.5em]

Expire, et inspire & Exhale, and inhale \\
Expire, et inspire & Exhale, and inhale \\
Expire\ldots & Exhale\ldots \\[1.2em]

\multicolumn{2}{l}{\textbf{Requiem Resonantiae} \hfill \textit{December 14, 2025}} \\
\multicolumn{2}{l}{\textit{Set to Rachmaninoff's Prelude in C-sharp minor.}} \\[0.3em]

Requiem aeternam & Eternal rest \\
et lux perpetua & and perpetual light \\
Requiem aeternam & Eternal rest \\
et resonantia & and resonance \\[0.5em]

Libera animas & Free the souls \\
ne absorbeat eas tartarus, & let not Tartarus absorb them, \\
Libera animas & Free the souls \\
ne cadant in obscurum. & let them not fall into darkness. \\[0.5em]

Quando caeli movendi sunt et terra & When the heavens and the earth are to be moved \\
in paradisum & into paradise \\[0.5em]

Libera me (Resonantia) & Free me (Resonance) \\
Libera te (Resonantia) & Free yourself (Resonance) \\[1.2em]

\pagebreak

\multicolumn{2}{l}{\textbf{Je me permets} \hfill \textit{December 3, 2025}} \\
\multicolumn{2}{l}{\textit{Set to the middle section of Rachmaninoff's Prelude in C-sharp minor.}} \\[0.3em]

Je me permets & I allow myself \\
Je me permets & I allow myself \\
de me concerner : & to concern myself with this: \\
les doux cernes sous tes yeux & the soft circles beneath your eyes \\[0.5em]

Je me permets & I allow myself \\
de discerner & to discern \\
les dissonances & the dissonances \\
de tes énoncés & in your utterances \\[0.5em]

Je me permets & I allow myself \\
de m'insérer & to insert myself \\
Je me permets & I allow myself \\
de te serrer & to hold you close \\
Je me permets & I allow myself \\
de te saisir & to grasp you \\
Je me permets\ldots & I allow myself\ldots \\
Ton désir & Your desire \\
Bien sûr, tout imaginaire. & Of course, entirely imagined. \\[0.5em]

Je me permets & I allow myself \\
de te cibler & to single you out \\
Tes forces et tes faiblesses & your strengths and your weaknesses \\
Rien n'est sûr à part les blessures & Nothing is certain except the wounds \\[0.5em]

Ça sert à qui, toutes ces folies & Whom do all these follies serve, \\
Qui animent mes nuits & that animate my nights \\
Ça sert à quoi, toutes ces histoires & What are all these stories for, \\
Rien n'est jamais acquis à l'homme & Nothing is ever fully acquired by man \\
Un môme & A child \\[0.5em]

Je me permets & I allow myself \\
Te connaître & to know you \\
Tu te permets ? & Do you allow yourself? \\
(Te renaître ?) & (To be reborn?) \\[1.2em]

\multicolumn{2}{l}{\textbf{Amis anéantis} \hfill \textit{December 30, 2025}} \\
\multicolumn{2}{l}{\textit{Set to the Prelude in C minor from Bach's \emph{Well-Tempered Clavier}.}} \\[0.3em]

Les amis anéantis & The undone friends \\
L'amour sans la cour & Love without courtship \\
L'amitiéternité & Friendsh-eternity \\
Oublier l'obligé & Forget the obliged \\[0.5em]

Observer au-delà & Observe beyond \\
Contempler, compte pas & Contemplate, do not count \\
Passer et repasser & Pass and pass again \\
Former et transformer & Form and transform \\[0.5em]

Accepter pas l'accès & Accept, not access \\
Accéder pas l'excès & Access, not excess \\
Saigner et sublimer & Bleed and sublimate \\
Soigner et s'éloigner & Heal and move away \\[0.5em]

Tenir et retenir & Hold and hold back \\
Contenter contenir & Be content to contain \\
Soulager les soupirs & Relieve the sighs \\
Rappeler : respirer & Remind: breathe \\
Expire, inspire, & Exhale, inhale, \\
Respire, tu m'inspires & Breathe, you inspire me \\[0.5em]

Tu cherches, chéri, & You search, dear, \\
Rigidité, & rigidity, \\
Dirige, diffère, & direct, defer, \\
déchire désir. & tear desire apart. \\
Délie, déplie, & untie, unfold, \\
détends, c'est temps & relax, it is time \\[0.5em]

Masser la gravité & Massage gravity \\
Graver la témérité & Engrave recklessness \\
Témoigner pas mériter & Bear witness, not deserve \\
La Jouissance & Jouissance \\[0.5em]

L'impasse, ça passe. & The impasse, it passes. \\
patience. L'Alliance & patience. The Alliance \\
des âmes incarnées, & of incarnated souls, \\
des amis, des amis anéantis & of friends, of devastated friends \\

\end{longtable}

\end{document}